\documentclass[12pt,preprint]{aastex}
\usepackage{epsfig,emulateapj5,apjfonts,aastexug}

\def\pn{\par\noindent}
\def\chandra{{\it Chandra}}

\def\gsimeq{\hbox{\raise0.5ex\hbox{$>\lower1.06ex\hbox{$\kern-1.07em{\sim}$}$}}} 
\def\lsimeq{\hbox{\raise0.5ex\hbox{$<\lower1.06ex\hbox{$\kern-1.07em{\sim}$}$}}} 

\shorttitle{Stacked EROs in the CDF--S}
\shortauthors{Brusa et al. }

\begin{document}


\title{Extremely Red Objects: an X--ray dichotomy}


\author{M. Brusa$^{1,2}$,A. Comastri$^2$, E. Daddi$^3$, A. Cimatti$^4$, M. Mignoli$^2$, L. Pozzetti$^2$} 
\affil{$^1$ Dipartimento di Astronomia, Universit\`a di Bologna,
via Ranzani 1, I-40127 Bologna, Italy}
\affil{$^2$ INAF --  Osservatorio Astronomico di Bologna, via Ranzani 1, 
I--40127 Bologna, Italy}
\affil{$^3$ European Southern Observatory, Karl--Schwarzschild--Strasse 2,
D--85748 Garching, Germany}
\affil{$^4$ INAF -- Osservatorio Astrofisico di Arcetri, Largo E. Fermi 5, 
I--55025 Firenze, Italy}


\begin{abstract}
\pn
We analyze the X--ray properties of a near--infrared selected ($K_s < 20$) 
sample of spectroscopically identified Extremely Red Objects ($R-K_s > 5$) 
in a region of the {\it Chandra} Deep Field--South using the public 1 Ms
observation. The 21 objects were classified, on the basis of
deep VLT spectra, in two categories: 13 dusty star--forming galaxies  
showing [OII] emission, and 8 early--type galaxies with
absorption features in their optical spectra. 
Only one emission line object has been individually detected;
its very hard X--ray spectrum and the high intrinsic X--ray luminosity 
unambiguously reveal the presence of an obscured AGN.
Stacking analysis of the remainder 12 emission line objects 
shows a significant detection with an average luminosity 
$L_X\approx8\times10^{40}$ erg s$^{-1}$ in the rest--frame 2--10 keV band.
The stacked counts of the 8 passive galaxies do not provide 
a positive detection.
We briefly discuss the implications of the present results 
for the estimate of the Star Formation Rate (SFR) in emission line
EROs.
\end{abstract}



\section{Introduction}
Having the colors expected for high-redshift  passive
ellipticals, Extremely Red Objects (hereinafter EROs; $R-K>5$) 
can be used as tracers of distant and old spheroids 
as a test for different cosmological models.
The optical and near--infrared properties of EROs are also consistent with 
those expected for high--redshift dusty starbursts
(e.g., HR 10, Cimatti et al. 1998; Smail et al. 1999) and Active 
Galactic Nuclei (AGN) reddened by dust (Pierre et al. 2001),
for which the spectral energy distribution at these wavelengths 
is dominated by the host galaxy starlight (e.g., Mainieri et
al. 2002).\\
The relative fraction of old ellipticals, 
dusty star--forming systems and obscured AGN among optically selected 
ERO samples is a key parameter in the study of the galaxy evolution
and can constrain models which link the formation of 
massive elliptical galaxies and the onset of AGN activity (Almaini et
al. 2002; Granato et al. 2001).
A step forward in this direction has been recently achieved by
the {\tt $K20$ survey} (Cimatti et al. 2002a; Daddi et
al. 2002). 
Deep VLT spectroscopy of a bright subsample at $K_s<19.2$ 
has shown that EROs are nearly equally populated by old, passively 
evolving systems and dusty star--forming galaxies  
over a similar range of redshift ($z_{mean}\simeq1$ for both the classes).
\pn
Sensitive hard X--ray observations provide a unique tool to uncover 
hidden AGN activity among the EROs population. 
Indeed, a sizeable fraction (10--30\%) of hard X-ray sources 
recently discovered in deep Chandra and XMM surveys are associated 
with EROs and the fraction depends on the limiting fluxes reached both
in the optical and X-ray bands (Alexander et al. 2002; 
Mainieri et al. 2002; Crawford et al. 2001).
The hard X--ray spectra 
indicate that X--ray bright EROs are associated with 
high redshift, highly obscured AGNs.\\ 
Moreover, thanks to the very deep limiting fluxes reached by 
deep {\it Chandra} surveys (CDFS, Giacconi et al. 2002, hereinafter G02; 
CDFN, Brandt et al. 2001a) it is also possible to investigate via stacking 
analysis the average X--ray properties of fainter non--AGN sources 
(Brandt et al. 2001b; Hornschemeier et al. 2002; Nandra et al. 2002). 
\pn
On the other hand, the AGN fraction in optically selected EROs samples
is still very uncertain and possibly lower than 15\% 
(Brusa et al. 2002; 
Alexander et al. 2002), 
suggesting that the bulk of the EROs population is not related to active
phenomena.
In fact, the majority of optically selected EROs
remains undetected even at the deepest limiting fluxes 
of the 1 Ms {\it Chandra} Deep Field North observation
(Alexander et al. 2002). Stacking analysis of those objects 
not individually detected suggests a softer X--ray spectrum 
than that of AGN--powered EROs and an X--ray to optical flux 
ratio 
consistent with that expected from 
normal elliptical galaxies, star--forming systems or 
low luminosity ($L_{2-10 keV} < 10^{42}$ erg s$^{-1}$)
nuclear activity at $z \simeq$ 1.
\pn
Here we make use of the available spectroscopic information of 
a well--defined, magnitude limited ($K_s < 20$) subsample of EROs
in the {\it Chandra} Deep Field--South area obtained as a part of the
{\tt K20 survey} (Cimatti et al. 2002a); 
the main aim of the present study is to investigate 
whether the X--ray emission properties of EROs 
are depending on the optical/infrared spectral properties.\\
Throughout the paper, a cosmology with $H_0=70$ km s$^{-1}$ Mpc$^{-1}$,
$\Omega_m$=0.3 and $\Omega_{\Lambda}$=0.7 is adopted. 

\section{The sample}
\pn
The ESO VLT {\tt K20} sample is one of the largest and most complete
spectroscopic sample of galaxies with $K_s<20$ available to date; 
the main scientific goal of this project is the comparison
of the observed redshift distribution 
of a complete sample of about 500 galaxies 
(87\% complete to $K_s<20$, 98\% if photometric redshifts are included)
with the predictions of galaxy
formation models in order to obtain stringent clues on the formation 
and evolution of the present-day massive galaxies (Cimatti et al. 2002b). 
The {\tt K20} sample cover a sub-area (32.2
arcmin$^2$) of the {\it Chandra} Deep Field South (CDFS) and a
19.8 arcmin$^2$ field around the quasar 0055-2659.
\pn
Extremely Red Objects in the {\tt K20} sample were selected 
with the criterion of $R-K_s>5$, and then classified 
on the basis of their optical spectra into two categories
(Cimatti et al. 2002a): 
objects with an absorption line spectrum, 
consistent with that of early--type, passively evolving galaxies 
(hereinafter ``{\tt old}'') 
and objects showing [OII]$\lambda3727$ emission line 
with an observed Equivalent Width EW$>$ 20 \AA over a red continuum, 
consistent with that of dusty star--forming systems (hereinafter
``{\tt dusty}''). 
The K20 EROs sample in the CDFS area includes 48 objects at $K_s<20$. 
For the purposes of the present paper we consider only the 
21 spectroscopically identified EROs which have been 
classified in the two categories described above:
13  ``dusty''  and 8 ``old'', distributed on a similar
range of redshifts, $z=0.8\div1.6$.

\section{X--ray data analysis}
\pn
The {\it Chandra} Deep Field South (CDFS) survey is a combination of
11 individual {\it Chandra} ACIS--I pointings 
observed in two years, from March 1999 to December 2000 (G02).
The maximum effective exposure time is 942 ks, 
but it varies accross the detector due to the CCD gaps and the
rotation of the field--of--view in the different observations.
In order to estimate the effective exposure time for each 
object, we have downloaded from the archive, along with the combined 
photon event list, all the events corrisponding to the single pointings.
We have calculated the exposure and instrument maps using standard
{\tt CIAO 2.2.1} tools for each observation separately, and combined
them to evaluate point--by--point the effective exposures.
As the mirror vignetting is energy dependent, we calculated the exposure maps 
at two energies (1 keV and 4 keV), representative of the mean energy 
of the photons in the soft (0.5--2 keV) and hard (2--8 keV) bands.\\
At first, we have searched for 
nearby 
(within $2\arcsec$) X-ray sources found by
{\sc wavdetect} in the K20 region, 
with a false-positive probability threshold of $10^{-5}$
(Freeman et~al. 2001).
Only one source, a dusty ERO at $z=1.327$, was 
individually detected; 
the optical position is almost coincident ($\Delta <0.3"$, within the
accuracy of the K--band position) 
with the {\it Chandra} source CXO CDFS~J033213.9-274526
in the G02 catalog.
The X--ray and optical properties of this object 
are discussed in Sect. 4.1.
\pn
In order to constrain average X--ray properties of the remaining
individually undetected EROs, we have applied the 
stacking technique (Brandt et al. 2001b; Nandra et al. 2002)  
separately for the two classes of objects. The samples 
consist of 12 ``dusty'' and 8 ``old'' EROs,  
with average redshifts of $z=1.053$  and $z=1.145$, respectively.
\pn
For the photometry we used a circular
aperture with a radius of $2''$ centered at the positions of the
selected EROs.  
We have also checked different aperture radii ($1-4"$) 
and found that the adopted one is an optimal extraction radius 
for the counts estimate, in agreement with the Nandra et al. (2002)
findings. 
The counts were stacked in the standard soft, 
hard and full bands (0.5--2 keV, 2--8 keV, and 0.5--8 keV),
and in the 1--5 keV band which roughly corresponds to the rest--frame
2--10 keV band at the mean redshift of the objects. 
\pn
To verify whether the stacked net counts constitute a significant signal 
the local background has been estimated: in annular regions around
each ERO centroid, as the average of eigth circular regions (2 arcsec
radius) for each ERO position and from a background map produced by
the WAVDETECT algorithm. The different choiches do not  significantly
affect the results in none of the adopted energy bands, except the
2--8 keV range where the results are strongly affected by the
increased level of the instrumental background.
Extensive Monte Carlo simulations (up to 100,000 trials) 
have been also carried out shuffling 12(8) random positions 
for ``dusty'' and ``old'' EROs respectively, using the same photometry 
aperture (2 arcsec). The random positions were chosen to lie in 
``local background regions''  to reproduce the actual background 
as close as possible. The resulting distributions are well approximated by 
Gaussians. The corresponding detection confidence level is in all the
cases in good 
agreement with that computed from Poisson statistic.
\pn
The results of the stacking analysis reported in Table 1 are referred 
to the background estimated as the average of the eight circular regions;
in Fig.~1 we show the summed images in the full (0.5--8 keV) band.
An excess of counts above the expected background level is clearly 
present for the 12 ``dusty'' EROS. 
\pn
The signal is stronger in the full band (99.997\% confidence level
assuming a Poissonian distribution, corresponding to about
4.2$\sigma$) and it is still present,  
although at a slightly lower confidence level, in the soft and 
1--5 keV band, while it is not statistically significant in the hard band.
Assuming a $\Gamma = 2$ power law spectrum plus Galactic absorption 
(N$_H=8\times 10^{19}$ cm$^{-2}$), 
the 3$\sigma$ upper limit on the average band ratio 
(H/S$<$0.85, where H and S indicates the hard band 
and soft band counts, respectively),  
corresponds to an absorption column density lower than $10^{22}$ cm$^{-2}$
(see Fig. 4 in Alexander et al. 2002). 

Given that the X--ray spectrum of nearby starburst galaxies is
well--represented by power--law models with slopes in the range
1.8$\div 2.5$ and low intrinsic absorption (e.g., Ptak et
al. 1999; Ranalli et al. 2002a), we adopted $\Gamma$=2.1 to compute
X--ray fluxes and luminosities. 
With this assumption, the stacked count rate in the 1--5 keV band
corresponds to an average 2--10 keV rest--frame luminosity of  
$\sim 8 \times 10^{40}$~erg~s$^{-1}$, at the mean redshift z=1.053. 
\pn
The X--ray emission from ``old'' EROs remains undetected 
in all the considered energy bands (see Table 1 for details). 
Assuming a thermal emission model with kT=1 keV and solar metallicity,
fully consistent with the average X--ray spectrum of nearby elliptical
galaxies (e.g., Pellegrini 1999), the 3$\sigma$ upper limit 
on the 0.5--2 keV luminosity is $L_X < 10^{41}$ erg s$^{-1}$
(see also Vignali et al. 2002).
\figurenum{1}
\centerline{\includegraphics[width=8.5cm]{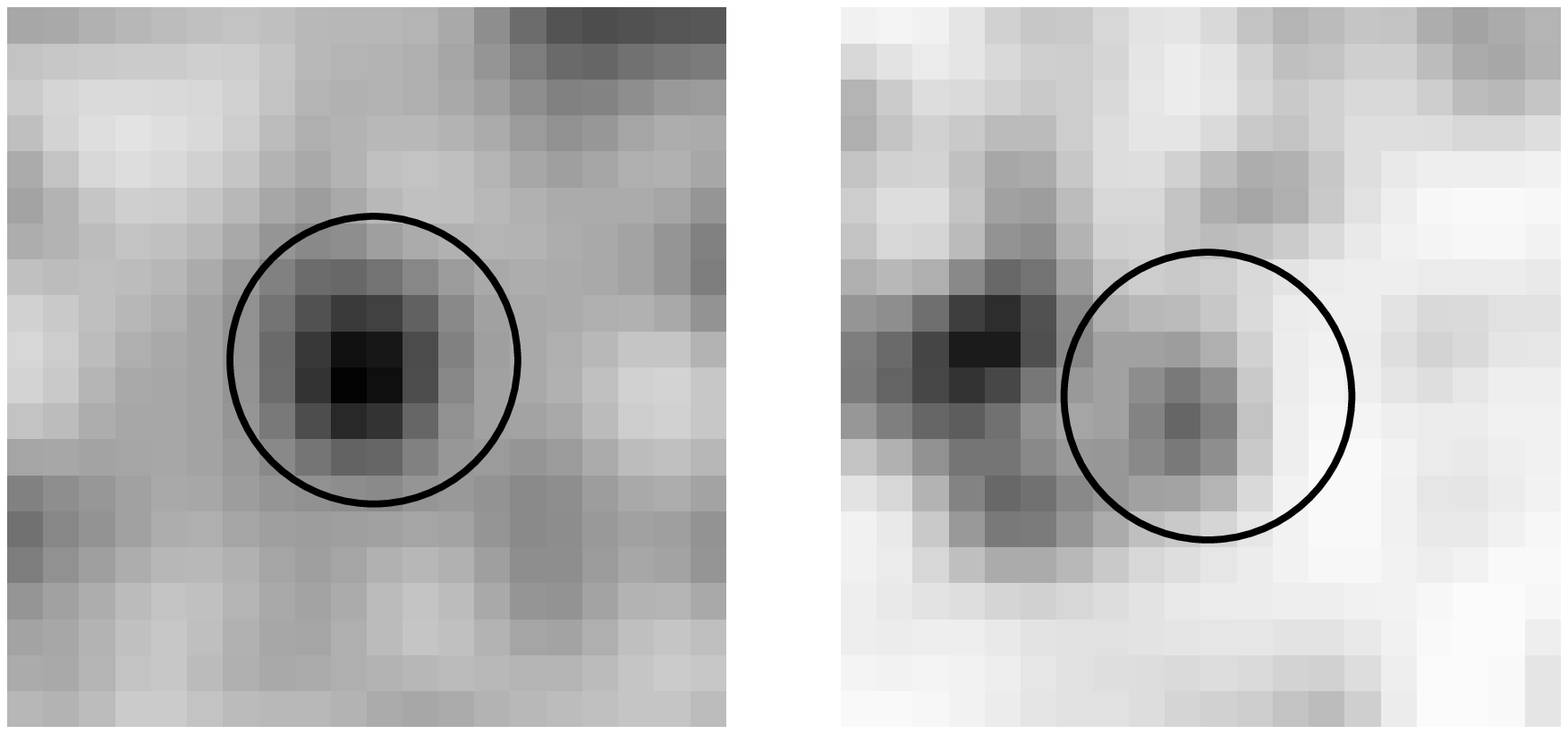}}
\vskip -1cm \figcaption{\footnotesize 
Stacked full--band images of the ``dusty'' ({\it left
panel}) and ``old'' ({\it right panel}) EROs. 
The images are 9x9 arcsec and have been smoothed with a gaussian of 
$\sigma$=1.5 pixel (approx 0.7$''$) . 
The circles are centered on the stacking position and have 
a radius of 2$''$. The detection significance of the summed counts is
4.2$\sigma$ for the ``dusty'' and $<2\sigma$ for the ``old''. 
The bright spot in the right panel at about 4$''$ from the stacked
position, corresponds to an individually detected object. 
\centerline{}
\label{fig1}}
%
%
\scriptsize
\begin{center}
{\sc TABLE~1 \\ Stacking Results}
\vskip 4pt
\begin{tabular}{lccccccc}
\hline
\hline
Class & band & counts & counts & c.l. & Exp. & 
Flux$^{\dag}$ & L$_X^{\dag}$ \\
  & (keV) & total & BKG &  & (Ms) & & \\
\hline
\small Dusty & 0.5-8 & 
	       172 &  124.2 & 
	       99.997\% & 
	       10.2 & 
	       3.4$\pm 1.3$ & 
	       $20 \pm 8$ \\ 
             & 0.5-2 & 
	       	67 & 37.9 & 
	       99.995\% & 
		10.4 & 
		1.3$\pm 0.45$ & 
		$7.4 \pm 2.6$ \\ 
             & 1-5  & 
		83 & 61.4 & 
		99.6\% & 
		10.2 &  
		1.4$\pm 0.5$ & 
		$8.0 \pm 2.8$ \\ 
             & 2-8  & 
		105 & 86.3 & 
		97.2\% & 
		10.1 & 
		$<4.8$ & 
		$ <27$ \\ 
\hline
\small Old$^{\ddag}$   & 0.5-8 & 
		102  & 92.1 & 
		$<$ 90\% &
		7.0 & 
		$<7.5$ & 
		$< 40$ \\ 
	     & 0.5-2 & 
		33 & 26.0  & 
		$<90$\% & 
		7.0 &
		$<24$ & 
                $< 10$ \\ 
\hline
\end{tabular}
\vskip 2pt
\end{center}
{\baselineskip 9pt
\footnotesize
\indent
$^{\dag}$: Fluxes (in units of 10$^{-17}$ erg cm$^{-2}$ s$^{-1}$) 
and luminosity (in units of 10$^{40}$ erg s$^{-1}$) are
computed assuming a power--law spectrum with 
$\Gamma=2.1$ and Galactic absorption for the ``dusty'' sample, and a
thermal emission model with kT=1 keV and solar metallicity for the 
``old'' sample. Errors are at the 90\% confidence level. \\
$^{\ddag}$: For the ``old'' sample we report only the results in the
0.5--8 keV and 0.5--2 keV bands. In the harder bands the results are
always consistent, within 1$\sigma$, with the background fluctuations.
}
\vglue0.2cm
\label{tab1}
\normalsize
%
\section{Results}

\subsection{CXO CDFS~J033213.9-274526: a type 2 AGN} 

The only ERO individually detected in the 1 Ms exposure
is the highest redshift, highest luminous 
(R=23.81, $K_s$=18.44) object in the ``dusty'' sample.
It is a very hard X--ray source 
(F$_{2-8 keV}$=$7.34\times 10^{-16}$ erg cm$^{-2}$ s$^{-1}$, 
$\Gamma=1.4$ and N$_H$=Gal; G02) undetected in the soft band 
(F$_{0.5-2 keV} < 5.48\times 10^{-17}$ erg cm$^{-2}$)
The observed band ratio (H/S$>$ 7.3) implies an intrinsic X--ray
column density larger than $4\times 10^{23}$ cm$^{-2}$, for a
classic AGN power--law spectrum with $\Gamma=1.8$).
The observed 2--8 keV luminosity is $\sim 7\times 10^{42}$ erg
s$^{-1}$ which corresponds to an unabsorbed, rest--frame hard X-ray
luminosity $\gsimeq4\times 10^{43}$ erg s$^{-1}$, typical of Seyfert 
galaxies. 
At the high--energy, therefore, CXO CDFS~J033213.9-274526 is naturally
classified as a Type 2, active object.
In the optical domain it shows only a faint, but significant,
[OII]$\lambda3727$ emission feature. Unfortunately, at the expected 
[NeV]$\lambda3426$ position a strong cosmic ray has affected the data.
The lack of (broad) MgII$\lambda2798$ emission indicates 
that obscured activity is taking place also in the optical nucleus.\\
Given the small area sampled and the spectroscopic completeness,
the present ERO sample is not suitable for a reliable estimate of
the AGN fraction among the EROs population. However, the present
low detection rate of AGN among EROs (1/21) confirms that the bulk
of the EROs population is not related to active phenomena.

\subsection{Dusty EROs}
The X--ray and optical data both suggest that starburst activity,
rather than AGN, is powering ``dusty'' EROs.
Indeed, following Cimatti et al. (2002a), we  recomputed the 
average spectrum of the 12 EROs we used in the X--ray stacking analysis.
The average optical spectrum is in excellent agreement with that 
published in Cimatti et al. 2002a (see their Fig. 2), derived from
the complete K20 EROs sample, except for the presence of the AGN indicator
[NeV]$\lambda3426$; in the present ``dusty'' sample this feature can be safely
ruled out (EW$<$1.5 \AA\ at the 3$\sigma$ level),   
suggesting that the optical emission is dominated by star--forming
systems.\\
The inferred, average X--ray luminosity 
is around one order of magnitude larger than that of normal spiral 
galaxies (Matsumoto et al. 1997) and it is similar to that of the 
starburst galaxy M82 (e.g., Griffiths et~al. 2000; Kaaret et~al. 2001);
the average X--ray spectrum (see Sect. 3) 
is consistent with that measured for nearby star--forming galaxies
(e.g. Dahlem, Weaver \& Heckman 1998; Ptak et al. 1999).\\ 
We have also investigated whether the stacked signal 
could be due to a few individual sources just below the detection 
threshold. This is particulary relevant
for our sample which may contain 
a few sub-threshold AGN responsible for the entire stacked signal. 
Five out of 12 ``dusty'' EROs are individually 
``detected'' at the 95--99\% confidence level.
The X--ray luminosity for these ``marginally detected'' EROs 
never exceeds $6\times~10^{41}$ erg s$^{-1}$ in the full--band  
excluding the presence of a bright AGN 
among the stacked dusty EROs.
Although the presence of a low--luminosity AGN (LLAGN) can not
be completely ruled out on the basis of the present analysis,  
in the following we assume that the ``dusty'' EROs are mainly powered by 
starburst related processes.

\subsubsection{Star Formation Rate: the hard X--rays view}

There is an increasing evidence that the hard X--ray luminosity of
star--forming objects can be used as a star formation rate (SFR)
indicator (Ranalli, Comastri, \& Setti 2002a,b; Nandra et al. 2002; 
Bauer et al. 2002).\\
Using the Ranalli et al. relation\footnote{SFR (M$_{\sun}$ yr$^{-1}$) =2.0
			      $\times 10^{-40}$ L$_{2-10 keV}$ (erg s$^{-1}$)}
we have derived an average SFR for the EROs in the ``dusty'' sample
of 5--44 M$_{\sun}$ yr$^{-1}$, 
taking into account the observed dispersion between the far--infrared
and X--ray luminosities quoted by Ranalli et al., and the
uncertainties on the hard X--ray luminosity. 
%
The average SFR derived from the the observed [OII] luminosity (Kennicutt
1998) is SFR([OII])$_{obs}$=2.7 M$_\sun$ yr$^{-1}$. 
The discrepancy between the optical and X--ray SFR estimates 
confirms that the optical emission lines are severely obscured by
dust and the optical value can be addressed as a {\it lower limit}. 
In order to provide a dereddened SFR estimate, the average spectrum 
was compared with template spectra of star forming galaxies and dusty
systems (as extensively discussed in Cimatti et al. 2002a); the best--fit
estimates for the reddening are in the range $E(B-V)=0.5\div0.8$, which
implies an intrinsic average SFR of the order 37--110 M$_{\sun}$ yr$^{-1}$
according to the Kennicutt (1998) relation.
Although the two estimates suffer of large uncertainties,
mainly related to the scatter of the
various SFR indicators, they can be reconciled if the lowest extinction
correction (e.g., $E(B-V)\sim0.5$, or even lower) is applied 
to correct the [OII] luminosity.
\par\noindent
From the derived X--ray SFR it is possible to 
estimate the average far--infrared (60--100 $\mu$m) and radio (1.4
GHz) emission  of the objects in our sample: 
L$_{FIR}=10^{10.4}\div10^{11.4}$ L$_{\sun}$; L$_{1.4 GHz}=2\div 17.6 \times
10^{29}$ erg s$^{-1}$ Hz$^{-1}$ (Kennicutt 1998).  
The relatively low value of L$_{FIR}$ would naturally explain the
low detection rate of EROs in submillimeter continuum
follow--up observations, typically sensitive to detect Ultra--Luminous
Infrared Galaxies (ULIGs, L$>10^{12}$ L$_{\sun}$) at $z\gsimeq 1$ 
(e.g., Mohan et al. 2002). 
The ``dusty'' ERO population would then be representative of 
less luminous dusty star--forming systems, with respect 
to submillimeter--selected high--z galaxies, as already pointed out in
Cimatti et al (2002a).\\ 
The inferred radio luminosity corresponds to K--corrected 
(S$_{\nu}\propto \nu^{-\alpha}$, $\alpha=0.7$) radio fluxes in 
the range 4-38$\mu$Jy. The lack of deep radio observations 
in the {\it Chandra} Deep Field South area prevents to test 
these predictions; however, 
it is interesting to note that very deep 1.4 GHz VLA observations
of a sample of optically selected EROs (Smail et al. 2002) agree
with our estimates (about 85\% of the EROs in their sample is not
detected  in the radio band at fluxes higher than $\sim 40$ $\mu$Jy). 

\section{Summary}
Thanks to the deep limiting fluxes reached in the 
1 Ms exposure of the {\it Chandra} Deep Field South
and to the subarcsec positional accuracy of the X--ray satellite,
it was possible to constrain via stacking analysis 
the high--energy properties of a well--defined,
spectroscopically identified sample of EROs, belonging to the {\tt K20 
survey}.  \\
The dichotomy in the optical classifications of EROs seems to be 
present also in the X--ray band: 
   ``dusty'' objects are relatively bright X--ray sources, 
   while ``old'' EROs are below the detection threshold.\\
The only object (CXO CDFS~J033213.9-274526) individually
   detected by {\it Chandra} is a luminous, 
   highly absorbed AGN (N$_H>4\times10^{23}$ cm$^{-2}$) at $z=1.327$.\\ 
Using the hard X--ray luminosity as a SFR indicator,
   which is essentially free from intrinsic absorption,  
   we have estimated a relatively low SFR (5--44 M$_{\sun}$ yr$^{-1}$)
   with respect to the estimate based on the reddening--dependent
   [OII] emission; our estimate is consistent with the results of recent
   far--infrared and radio observations.\\
Deep X--ray observations of large and complete samples of spectroscopically
identified EROs could provide a powerful tool to estimate, via
stacking analysis, the average star formation rate in ``dusty''
systems and to better assess the AGN fraction among the ERO
population.

\acknowledgments
The authors acknowledge partial support by ASI I/R/103/00 and I/R/113/01
contracts and MIUR Cofin--00--02--36 grant.   
MB and AC thank P. Ranalli, C. Vignali and G. Zamorani
for useful comments and discussions. We thank the anonymous referee
for his/her helpful comments.

\end{document}